\documentclass[12pt]{article}
\usepackage{amsmath,amssymb,amsthm,amsxtra,overpic,bbm,bm,epsfig}
\usepackage{color}
=-24mm

\usepackage{hyperref}
\usepackage{url}
\usepackage{slashed,stmaryrd}
\def\thefootnote{\fnsymbol{footnote}}
\vspace{-3cm}
\title{\Large\bf Commutators of lepton mass matrices associated with \\
seesaw and leptogenesis}
\author{\normalsize {\bf Yikun Wang}$^a$, {\bf Zhi-zhong Xing}$^{b,a,c}$
\footnote{E-mail: xingzz@ihep.ac.cn} \\ \vspace{-0.3cm} \\
\small $^a$SYSU-IHEP School for High Energy Physics, Sun Yat-Sen
University, Guangzhou 510275; \\
\small $^b$Institute of High Energy Physics, Chinese Academy of
Sciences, Beijing 100049; \\
\small $^c$Center for High Energy Physics, Peking University,
Beijing 100080, China}

\begin{document}

\maketitle

\begin{abstract}
The origin of tiny neutrino masses and the baryon number asymmetry
of the Universe are naturally interpreted by the canonical seesaw
and leptogenesis mechanisms, in which there are the heavy Majorana
neutrino mass matrix $M^{}_{\rm R}$, the Dirac neutrino mass matrix
$M^{}_{\rm D}$, the charged-lepton mass matrix $M^{}_\ell$ and the
effective (light) neutrino mass matrix $M^{}_\nu$. We find that
${\rm Im}\left(\det\left[ M^\dagger_{\rm D} M^{}_{\rm D},
M^\dagger_{\rm R} M^{}_{\rm R} \right]\right)$, ${\rm
Im}\left(\det\left[ M^{}_\ell M^\dagger_\ell, M^{}_\nu M^\dagger_\nu
\right]\right)$ and ${\rm Im}\left(\det\left[ M^{}_\ell
M^\dagger_\ell, M^{}_{\rm D} M^\dagger_{\rm D} \right]\right)$ can
serve for a basis-independent measure of CP violation associated
with lepton-number-violating decays of heavy neutrinos, flavor
oscillations of light neutrinos and lepton-flavor-violating decays
of charged leptons, respectively. We first calculate these
quantities with the help of a standard parametrization of the
$6\times 6$ flavor mixing matrix, and then discuss their
implications on both leptogenesis and CP violation at low energy
scales. A comparison with the weak-basis invariants of leptogenesis
as proposed by Branco {\it et al} is also made.
\end{abstract}

\begin{flushleft}
\hspace{0.8cm} PACS number(s): 14.60.Pq, 13.10.+q, 25.30.Pt \\
\hspace{0.8cm} Keywords: lepton flavor, neutrino mass, CP violation,
commutator, seesaw, leptogenesis
\end{flushleft}

\def\thefootnote{\arabic{footnote}}
\setcounter{footnote}{0}

\newpage

\framebox{\large\bf 1} \hspace{0.1cm}
The commutator of quark mass
matrices $M^{}_{\rm u}$ and $M^{}_{\rm d}$, or equivalently the
commutator of quark Yukawa coupling matrices $Y^{}_{\rm u}$ and
$Y^{}_{\rm d}$, has proved to be a quite useful measure of weak CP
violation in the standard model (SM) \cite{J85}. Given the quark
masses $m^{}_u \simeq 1.38$ MeV, $m^{}_d \simeq 2.82$ MeV, $m^{}_s
\simeq 57$ MeV, $m^{}_c \simeq 0.638$ GeV, $m^{}_b \simeq 2.86$ GeV
and $m^{}_t \simeq 172.1$ GeV at the electroweak energy scale $\mu =
M^{}_Z$ \cite{XZZ}, one may easily arrive at
\begin{eqnarray}
{\rm Im}\left(\det\left[Y^{}_{\rm u} Y^\dagger_{\rm u}, Y^{}_{\rm d}
Y^\dagger_{\rm d}\right]\right) \hspace{-0.2cm} & = &
\hspace{-0.2cm} \frac{2^7}{v^{12}} {\cal J}^{}_q \left(m^2_t -
m^2_u\right) \left(m^2_t - m^2_c\right) \left(m^2_c - m^2_u\right)
\left(m^2_b - m^2_d\right)
\left(m^2_b - m^2_s\right) \left(m^2_s - m^2_d\right) \nonumber \\
\hspace{-0.2cm} & \simeq & \hspace{-0.2cm} 6.0 \times 10^{-20} \; ,
%     (1)
\end{eqnarray}
where $v \simeq 246$ GeV denotes the vacuum expectation value of the
SM Higgs field, ${\cal J}^{}_q \simeq 2.96 \times 10^{-5}$ is the
Jarlskog invariant of CP violation in the quark sector \cite{PDG},
and the relation $M^{}_{\rm u,d} = Y^{}_{\rm u,d} v/\sqrt{2}$ has
been used. The very small number obtained in Eq. (1) is ten orders
of magnitude smaller than $\eta \equiv n^{}_{\rm b}/n^{}_\gamma
\simeq 6.2 \times 10^{-10}$, the observed baryon number asymmetry of
the Universe \cite{PDG}. This is one of the reasons why the SM
itself cannot account for the cosmological matter-antimatter
asymmetry.

A simple extension of the SM is to add three right-handed neutrinos
and allow lepton number violation in the lepton sector,
\begin{eqnarray}
-{\cal L}^{}_{\rm lepton} = \overline{\ell^{}_{\rm L}} Y^{}_\ell H
E^{}_{\rm R} + \overline{\ell^{}_{\rm L}} Y^{}_\nu \tilde{H}
N^{}_{\rm R} + \frac{1}{2} \overline{N^c_{\rm R}} M^{}_{\rm R}
N^{}_{\rm R} + {\rm h.c.} \; ,
%     (2)
\end{eqnarray}
where the notations for the SM fields are self-explanatory,
$N^{}_{\rm R}$ (or $N^c_{\rm R}$) is the column vector of the
right-handed neutrino fields (or its charge-conjugate counterpart),
and $M^{}_{\rm R}$ stands for the symmetric Majorana mass matrix.
The scale of $M^{}_{\rm R}$ is expected to be much larger than $v$,
because the right-handed neutrinos are $\rm SU(2)^{}_{\rm L}$
singlets. In this case the effective mass matrix of three light
neutrinos is approximately given by the seesaw relation \cite{SS1}
\begin{eqnarray}
M^{}_\nu \simeq -\frac{1}{2} Y^{}_\nu \frac{v^2}{M^{}_{\rm R}}
Y^T_\nu = - M^{}_{\rm D} \frac{1}{M^{}_{\rm R}} M^T_{\rm D} \; ,
%     (3)
\end{eqnarray}
where $M^{}_{\rm D} = Y^{}_\nu v/\sqrt{2}$ is usually referred to as
the Dirac neutrino mass matrix. The smallness of the mass scale of
$M^{}_\nu$ is therefore attributed to the largeness of the mass
scale of $M^{}_{\rm R}$. A special bonus of this canonical seesaw is
the effectiveness of the leptogenesis mechanism \cite{FY}, which
provides an elegant interpretation of $\eta \simeq 6.2 \times
10^{-10}$ for the observable Universe thanks to the
lepton-number-violating and CP-violating decays of heavy Majorana
neutrinos.

The present work aims to construct the commutators of lepton mass
matrices and explore their relations with CP violation in both
lepton-flavor-violating and lepton-number-violating processes. We
find that ${\rm Im}\left(\det\left[ M^\dagger_{\rm D} M^{}_{\rm D},
M^\dagger_{\rm R} M^{}_{\rm R} \right]\right)$, ${\rm
Im}\left(\det\left[ M^{}_\ell M^\dagger_\ell, M^{}_\nu M^\dagger_\nu
\right]\right)$ and ${\rm Im}\left(\det\left[ M^{}_\ell
M^\dagger_\ell, M^{}_{\rm D} M^\dagger_{\rm D} \right]\right)$ can
serve for a basis-independent measure of CP violation associated
with the lepton-number-violating decays of heavy neutrinos, the
flavor oscillations of light neutrinos and the
lepton-flavor-violating decays of charged leptons, respectively. We
calculate these rephasing-invariant quantities in terms of a
standard parametrization of the $6\times 6$ flavor mixing matrix,
and discuss their implications on both leptogenesis and CP violation
at low energy scales. For the sake of comparison, we also calculate
the weak-basis invariants of leptogenesis as defined by Branco {\it
et al} \cite{Branco} and point out their similarity with and difference
from our invariant ${\rm Im}\left(\det\left[ M^\dagger_{\rm D} M^{}_{\rm
D}, M^\dagger_{\rm R} M^{}_{\rm R} \right]\right)$.

\framebox{\large\bf 2} \hspace{0.1cm} After the $\rm SU(2)^{}_{\rm L}
\times U(1)^{}_{\rm Y}$ symmetry is spontaneously broken to
$\rm U(1)^{}_{\rm em}$, Eq. (2) becomes
\begin{eqnarray}
-{\cal L}^\prime_{\rm lepton} = \overline{E^{}_{\rm L}} M^{}_\ell
E^{}_{\rm R} + \frac{1}{2} ~ \overline{\left( \nu^{}_{\rm L}
~N^c_{\rm R}\right)} ~ \left( \begin{matrix} {\bf 0} & M^{}_{\rm D}
\cr M^T_{\rm D} & M^{}_{\rm R} \end{matrix}\right) \left(
\begin{matrix} \nu^c_{\rm L} \cr N^{}_{\rm R} \end{matrix}\right) +
{\rm h.c.} \; ,
%     (4)
\end{eqnarray}
where $M^{}_\ell = Y^{}_\ell v/\sqrt{2}$. Without loss of
generality, we choose a convenient lepton flavor basis in which
$M^{}_\ell = \widehat{M}^{}_\ell \equiv
{\rm Diag}\{m^{}_e, m^{}_\mu, m^{}_\tau\}$ holds. The
overall $6\times 6$ neutrino mass matrix is symmetric, and it can be
diagonalized by a unitary matrix containing 15 angles and 15
phases \cite{Xing12}:
\begin{eqnarray}
\left[\left( \begin{matrix} {\bf 1} & {\bf 0} \cr {\bf 0} & U^{}_0 \cr
\end{matrix} \right) \left( \begin{matrix}A & R \cr S & B \cr \end{matrix}
\right) \left( \begin{matrix} V^{}_0 & {\bf 0} \cr {\bf 0} & {\bf 1}
\cr \end{matrix} \right)\right]^\dagger \left( \begin{matrix} {\bf
0} & M^{}_{\rm D} \cr M^T_{\rm D} & M^{}_{\rm R} \end{matrix}\right)
\left[\left( \begin{matrix} {\bf 1} & {\bf 0} \cr {\bf 0} & U^{}_0
\cr \end{matrix} \right) \left( \begin{matrix}A & R \cr S & B \cr
\end{matrix} \right) \left( \begin{matrix} V^{}_0 & {\bf 0} \cr {\bf
0} & {\bf 1} \cr \end{matrix} \right)\right]^* = \left(
\begin{matrix} \widehat{M}^{}_\nu & {\bf 0} \cr {\bf 0} &
\widehat{M}^{}_N \end{matrix} \right) \; ,
%     (5)
\end{eqnarray}
where $\widehat{M}^{}_\nu \equiv {\rm Diag}\{m^{}_1, m^{}_2, m^{}_3
\}$ and $\widehat{M}^{}_N \equiv {\rm Diag}\{M^{}_1, M^{}_2, M^{}_3
\}$ with $m^{}_i$ or $M^{}_i$ (for $i=1,2,3$) being the physical
masses of light or heavy Majorana neutrinos, and $V^{}_0$ or
$U^{}_0$ is a $3\times 3$ unitary matrix which consists of three
mixing angles and three CP-violating phases. This basis
transformation allows us to express the weak charged-current
interactions of six neutrinos in terms of their mass states:
\begin{eqnarray}
-{\cal L}^{}_{\rm cc} = \frac{g}{\sqrt{2}} ~ \overline{\left(e~~
\mu~~ \tau\right)^{}_{\rm L}} ~ \gamma^\mu \left[ V \left(
\begin{matrix} \nu^{}_1 \cr \nu^{}_2 \cr \nu^{}_3 \end{matrix}
\right)^{}_{\rm L} + R \left( \begin{matrix} N^{}_1 \cr N^{}_2 \cr
N^{}_3 \end{matrix} \right)^{}_{\rm L} \right] W^-_\mu + {\rm h.c.} \; ,
%     (6)
\end{eqnarray}
where $V \equiv AV^{}_0$ is the Maki-Nakagawa-Sakata-Pontecorvo
(MNSP) matrix \cite{MNS} responsible for flavor oscillations of the
light neutrinos $\nu^{}_i$ (for $i=1,2,3$), and $R$ measures the
strength of charged-current interactions of the heavy neutrinos
$N^{}_i$ (for $i=1,2,3$). The relationship between $V$ and $R$ is
$VV^\dagger = AA^\dagger = {\bf 1} - RR^\dagger$ \cite{X2008}. Since
$R$ and $S$ describe the mixing between light and heavy neutrinos,
the magnitudes of their elements are constrained to be at most of
${\cal O}(0.1)$ \cite{Antusch}. The exact expressions of $A$, $B$,
$R$, $S$, $U^{}_0$ and $V^{}_0$ can be found in Ref. \cite{Xing12}.
Here we only quote
\begin{eqnarray}
V^{}_0 = \left( \begin{matrix} c^{}_{12} c^{}_{13} & \hat{s}^*_{12}
c^{}_{13} & \hat{s}^*_{13} \cr -\hat{s}^{}_{12} c^{}_{23} -
c^{}_{12} \hat{s}^{}_{13} \hat{s}^*_{23} & c^{}_{12} c^{}_{23} -
\hat{s}^*_{12} \hat{s}^{}_{13} \hat{s}^*_{23} & c^{}_{13}
\hat{s}^*_{23} \cr \hat{s}^{}_{12} \hat{s}^{}_{23} - c^{}_{12}
\hat{s}^{}_{13} c^{}_{23} & -c^{}_{12} \hat{s}^{}_{23} -
\hat{s}^*_{12} \hat{s}^{}_{13} c^{}_{23} & c^{}_{13} c^{}_{23} \cr
\end{matrix} \right) \; ,
%     (7)
\end{eqnarray}
where $c^{}_{ij} \equiv \cos\theta^{}_{ij}$, $\hat{s}^{}_{ij}
\equiv e^{{\rm i} \delta^{}_{ij}} s^{}_{ij}$ and
$s^{}_{ij} \equiv \sin\theta^{}_{ij}$ with $\theta^{}_{ij}$
and $\delta^{}_{ij}$ being angles and phases (for $1\leq i < j \leq
6$), respectively. In view of the smallness of the nine
active-sterile mixing angles $\theta^{}_{ij}$ (for $i=1,2,3$ and
$j=4,5,6$), we arrive at the following excellent approximations:
\begin{eqnarray}
A \simeq B \simeq {\bf 1} - {\cal O}(s^2_{ij}) \; , ~~~~ R \simeq
-S^\dagger \simeq \left( \begin{matrix} \hat{s}^*_{14} &
\hat{s}^*_{15} & \hat{s}^*_{16} \cr \hat{s}^*_{24} & \hat{s}^*_{25}
& \hat{s}^*_{26} \cr \hat{s}^*_{34} & \hat{s}^*_{35} &
\hat{s}^*_{36} \cr \end{matrix} \right) + {\cal O}(s^3_{ij}) \; .
%     (8)
\end{eqnarray}
In this case we have $M^{}_{\rm D} \simeq R\widehat{M}^{}_N U^T_0$,
$M^{}_{\rm R} \simeq U^{}_0 \widehat{M}^{}_N U^T_0$ and $V^{}_0
\widehat{M}^{}_\nu V^T_0 + R\widehat{M}^{}_N R^T \simeq {\bf 0}$.
The seesaw relation in Eq. (3) can therefore be reexpressed as
$M^{}_\nu \equiv V \widehat{M}^{}_\nu V^T \simeq V^{}_0
\widehat{M}^{}_\nu V^T_0 \simeq -R\widehat{M}^{}_N R^T$.

Let us first calculate the commutator $\left[M^\dagger_{\rm D}
M^{}_{\rm D}, M^\dagger_{\rm R} M^{}_{\rm R}\right]$, which is
essentially associated with CP violation in the
lepton-number-violating decays of $N^{}_i$ at very high energy
scales. Given the good approximations made in Eq. (8), the result is
\begin{eqnarray}
\hspace{-0.2cm} & & \hspace{-0.2cm} \left[ M^\dagger_{\rm D}
M^{}_{\rm D}, M^\dagger_{\rm R} M^{}_{\rm R} \right] \simeq U^*_0
\widehat{M}^{}_N \left[ R^\dagger R, \widehat{M}^2_N \right]
\widehat{M}^{}_N U^T_0
\nonumber \\
& \simeq & \hspace{-0.2cm} U^*_0 \widehat{M}^{}_N \left(
\begin{matrix} 0 & -M^{}_1 M^{}_2 \Delta^{}_{12}
\displaystyle\sum^3_{i=1} \hat{s}^{}_{i4} \hat{s}^{*}_{i5} & -M^{}_1
M^{}_3 \Delta^{}_{13} \displaystyle\sum^3_{i=1} \hat{s}^{}_{i4}
\hat{s}^{*}_{i6} \cr M^{}_1 M^{}_2 \Delta^{}_{12}
\displaystyle\sum^3_{i=1} \hat{s}^*_{i4} \hat{s}^{}_{i5} & 0 &
-M^{}_2 M^{}_3 \Delta^{}_{23} \displaystyle\sum^3_{i=1}
\hat{s}^{}_{i5} \hat{s}^{*}_{i6} \cr M^{}_1 M^{}_3 \Delta^{}_{13}
\displaystyle\sum^3_{i=1} \hat{s}^*_{i4} \hat{s}^{}_{i6} & M^{}_2
M^{}_3 \Delta^{}_{23} \displaystyle\sum^3_{i=1} \hat{s}^*_{i5}
\hat{s}^{}_{i6} & 0 \cr \end{matrix} \right) \widehat{M}^{}_N U^T_0
\; , \;\;\;\;\;\;
%     (9)
\end{eqnarray}
where $\Delta^{}_{ij} \equiv M^2_i - M^2_j$ (for $i,j=1,2,3$). The
determinant of this commutator is purely imaginary, and thus we
obtain
\begin{eqnarray}
{\rm Im}\left(\det\left[ M^\dagger_{\rm D} M^{}_{\rm D},
M^\dagger_{\rm R} M^{}_{\rm R} \right]\right) \simeq \det\left(
\widehat{M}^2_N\right) {\rm Im}\left(\det\left[ R^\dagger R,
\widehat{M}^2_N \right]\right) \simeq 2 M^2_1 M^2_2 M^2_3
\Delta^{}_{12} \Delta^{}_{13} \Delta^{}_{23} {\cal X}^{}_N \; ,
%     (10)
\end{eqnarray}
in which
\begin{eqnarray}
{\cal X}^{}_N \equiv
{\rm Im} \left[ \left(\sum^3_{i=1} \hat{s}^{}_{i4} \hat{s}^{*}_{i5}\right)
\left(\sum^3_{i=1} \hat{s}^{*}_{i4} \hat{s}^{}_{i6}\right)
\left(\sum^3_{i=1} \hat{s}^{}_{i5} \hat{s}^{*}_{i6}\right) \right] \;
%     (11)
\end{eqnarray}
is a Jarlskog-like quantity associated with the effects of
CP violation in the decays of heavy Majorana neutrinos. So
${\rm Im}\left(\det\left[ M^\dagger_{\rm D}
M^{}_{\rm D}, M^\dagger_{\rm R} M^{}_{\rm R} \right]\right)$
will vanish if $\Delta^{}_{ij} =0$ holds or if ${\cal X}^{}_N =0$ holds.
We see that ${\cal X}^{}_N$ depends on six independent
phase differences, such as $\delta^{}_{i 4} - \delta^{}_{i 5}$ and
$\delta^{}_{i5} - \delta^{}_{i 6}$ (for $i=1,2,3$). Of course, the
CP-violating asymmetries $\varepsilon^{}_{i\alpha}$
(or $\varepsilon^{}_i = \varepsilon^{}_{i e} + \varepsilon^{}_{i \mu} +
\varepsilon^{}_{i \tau}$ in the unflavored case) between the
lepton-number-violating decay modes $N^{}_i \to \ell^{}_\alpha + H$
and $N^{}_i \to \overline{\ell^{}_\alpha} + \overline{H}$ must
depend on the same phase differences \cite{Xing12}. However,
${\cal X}^{}_N \neq 0$ is in general a necessary but not sufficient condition
for $\varepsilon^{}_{i\alpha} \neq 0$ or $\varepsilon^{}_i \neq 0$,
and hence ${\cal X}^{}_N = 0$ cannot guarantee $\varepsilon^{}_{i\alpha} = 0$
or $\varepsilon^{}_i = 0$ (or vice versa) either.

If $M^{}_1 \ll M^{}_2 \ll M^{}_3$ holds and the leptogenesis
mechanism works at temperature $T \simeq M^{}_1$ \cite{FY}, then it
should not be difficult for the quantity
\begin{eqnarray}
\frac{{\rm Im}\left(\det\left[ M^\dagger_{\rm D} M^{}_{\rm D},
M^\dagger_{\rm R} M^{}_{\rm R} \right]\right)}{T^{12}} \simeq -2
\left(\frac{M^{}_2}{M^{}_1}\right)^4 \left(\frac{M^{}_3}{M^{}_1}
\right)^6 {\cal X}^{}_N \;\;\;
%     (12)
\end{eqnarray}
to be comparable with or larger than $\eta \simeq 6.2 \times
10^{-10}$ in magnitude. Taking $M^{}_3 \sim 10^2 M^{}_2 \sim 10^4
M^{}_1$ and $\theta^{}_{i 4} \sim \theta^{}_{i 5} \sim \theta^{}_{i
6} \lesssim {\cal O}(10^{-7})$ for example, we expect that $|{\cal X}^{}_N|
\lesssim {\cal O}(10^{-42})$ holds and thus the magnitude
of ${\rm Im}\left(\det\left[ M^\dagger_{\rm D} M^{}_{\rm D},
M^\dagger_{\rm R} M^{}_{\rm R} \right]\right)/T^{12}$ is in general
possible to reach the ${\cal O}(10^{-10})$ level. In fact, a number
of specific seesaw-plus-leptogenesis models have so far been
proposed to successfully account for the observed baryon number
asymmetry of the Universe \cite{BPY}.

We proceed to calculate the commutator $\left[M^{}_\ell M^\dagger_\ell,
M^{}_\nu M^\dagger_\nu \right]$, which is directly relevant to leptonic CP
violation in neutrino oscillations \cite{Xing2001}. Given the flavor
basis $M^{}_\ell = \widehat{M}^{}_\ell$ and
the good approximation $M^{}_\nu \simeq V^{}_0 \widehat{M}^{}_\nu
V^T_0$, it is straightforward to arrive at
\begin{eqnarray}
{\rm Im}\left(\det\left[ M^{}_\ell M^\dagger_\ell, M^{}_\nu
M^\dagger_\nu \right]\right) \simeq 2\Delta^{}_{e\mu}
\Delta^{}_{e\tau} \Delta^{}_{\mu\tau} \Delta^\prime_{12}
\Delta^\prime_{13} \Delta^\prime_{23} {\cal J}^{}_\nu \; , \;\;
%     (13)
\end{eqnarray}
in which $\Delta^{}_{\alpha\beta} \equiv m^2_\alpha - m^2_\beta$
(for $\alpha, \beta = e,\mu,\tau$) and $\Delta^\prime_{ij} \equiv
m^2_i - m^2_j$ (for $i,j = 1,2,3$) are defined, and ${\cal J}^{}_\nu
= c^{}_{12} s^{}_{12} c^2_{13} s^{}_{13} c^{}_{23}
s^{}_{23}\sin\delta$ with $\delta \equiv \delta^{}_{13}
-\delta^{}_{12} -\delta^{}_{23}$ is just the Jarlskog invariant of
$V^{}_0$. Note that the mixing angles and CP-violating phases of
$V^{}_0$ are actually correlated with those of $R$ due to the seesaw
relation $V^{}_0 \widehat{M}^{}_\nu V^T_0 + R \widehat{M}^{}_N R^T
\simeq {\bf 0}$. The latter allows us to express Eq. (13) in terms
of the parameters of $\widehat{M}^{}_N$ and $R$ besides
$\Delta^{}_{\alpha\beta}$, but the result is rather lengthy and thus
less instructive. A key point is that ${\cal X}^{}_N =0$ does not
necessarily lead to ${\cal J}^{}_\nu =0$, or vice versa,
as one can see from the seesaw formula. Similarly, it is possible
for either $\varepsilon^{}_{i\alpha} =0$ (or $\varepsilon^{}_{i} =0$)
but ${\cal J}^{}_\nu \neq 0$ or $\varepsilon^{}_{i\alpha} \neq 0$
(or $\varepsilon^{}_{i} \neq 0$) but ${\cal J}^{}_\nu = 0$ to hold.
Hence there is in general no direction connection between leptogenesis
and CP violation at low energy scales
%%%%%%%%%%%%%%%%%%%%%%%%%%%%%%
\footnote{However, it is possible to establish a direct
connection between leptogenesis and CP violation at low energy
scales in some specific seesaw models, in which most of the phase
parameters can be switched off \cite{BPY}.}.
%%%%%%%%%%%%%%%%%%%%%%%%%%%%%%

Let us estimate the magnitude of
${\rm Im}\left(\det\left[ M^{}_\ell M^\dagger_\ell, M^{}_\nu
M^\dagger_\nu \right]\right)$ in order to give one a ball-park
feeling of how small it is. The values of three charged-lepton masses
are $m^{}_e \simeq 0.48657$ MeV, $m^{}_\mu = 102.718$ MeV and
$m^{}_\tau \simeq 1746.17$ MeV at the electroweak
scale $\mu = M^{}_Z$ \cite{XZZ}. Furthermore, a global analysis
of current neutrino oscillation data
yields $\Delta^\prime_{12} \simeq
-7.5 \times 10^{-5} ~ {\rm eV}^2$, $\Delta^\prime_{13} \simeq
\Delta^\prime_{23} \simeq \mp 2.4 \times 10^{-3} ~ {\rm eV}^2$,
$\theta^{}_{12} \simeq 34^\circ$, $\theta^{}_{13} \simeq 9^\circ$,
$\theta^{}_{23} \simeq 41^\circ$ and $\delta \sim 250^\circ$
\cite{Fogli}. We therefore obtain
${\cal J}^{}_\nu \sim -3.3 \times 10^{-2}$ and
${\rm Im}\left(\det\left[ M^{}_\ell M^\dagger_\ell, M^{}_\nu
M^\dagger_\nu \right]\right) \sim -2.3 \times 10^{-64} ~{\rm GeV}^{12}$.
Because ${\cal J}^{}_\nu$ measures the strength of leptonic CP violation in
flavor oscillations of the light neutrinos, we anticipate some appreciable
CP-violating effects to occur in the forthcoming long-baseline experiments.

Analogous to the commutator $\left[ M^{}_\ell M^\dagger_\ell, M^{}_\nu
M^\dagger_\nu \right]$, the leptonic commutator
$\left[ M^{}_\ell M^\dagger_\ell, M^{}_{\rm R} M^\dagger_{\rm R} \right]$
can be used to describe CP violation
{\it within} the heavy Majorana neutrino sector. Given the
approximation $M^{}_{\rm R} \simeq U^{}_0 \widehat{M}^{}_N U^T_0$ and the
parametrization \cite{Xing12}
\begin{eqnarray}
U^{}_0 = \left( \begin{matrix} c^{}_{45} c^{}_{46} & \hat{s}^*_{45}
c^{}_{46} & \hat{s}^*_{46} \cr -\hat{s}^{}_{45} c^{}_{56} -
c^{}_{45} \hat{s}^{}_{46} \hat{s}^*_{56} & c^{}_{45} c^{}_{56} -
\hat{s}^*_{45} \hat{s}^{}_{46} \hat{s}^*_{56} & c^{}_{46}
\hat{s}^*_{56} \cr \hat{s}^{}_{45} \hat{s}^{}_{56} - c^{}_{45}
\hat{s}^{}_{46} c^{}_{56} & -c^{}_{45} \hat{s}^{}_{56} -
\hat{s}^*_{45} \hat{s}^{}_{46} c^{}_{56} & c^{}_{46} c^{}_{56} \cr
\end{matrix} \right) \; ,
%     (14)
\end{eqnarray}
It is easy to obtain
\begin{eqnarray}
{\rm Im}\left(\det\left[ M^{}_\ell M^\dagger_\ell, M^{}_{\rm R}
M^\dagger_{\rm R} \right]\right) \simeq 2\Delta^{}_{e\mu}
\Delta^{}_{e\tau} \Delta^{}_{\mu\tau} \Delta^{}_{12}
\Delta^{}_{13} \Delta^{}_{23} {\cal J}^{}_N \; , \;
%     (15)
\end{eqnarray}
where $\Delta^{}_{\alpha\beta}$ and $\Delta^{}_{ij}$ have already
been defined, and ${\cal J}^{}_N = c^{}_{45} s^{}_{45} c^2_{46}
s^{}_{46} c^{}_{56} s^{}_{56}\sin\delta^\prime$ with
$\delta^\prime \equiv \delta^{}_{46} -\delta^{}_{45} -\delta^{}_{56}$
denotes the Jarlskog invariant of $U^{}_0$. Comparing Eq. (14) with
Eq. (7), we see the exact parallelism between the parameters of
$U^{}_0$ and $V^{}_0$. It might be interesting to make a naive
conjecture: $\theta^{}_{45} = \theta^{}_{12}$,
$\theta^{}_{46} = \theta^{}_{13}$, $\theta^{}_{56} = \theta^{}_{23}$
and $\delta^\prime = \delta$, which in turn lead to
${\cal J}^{}_N = {\cal J}^{}_\nu$.
Whether such a heavy-light neutrino symmetry is phenomenologically useful
remains an open question.

Another interesting commutator is $\left[ M^{}_\ell M^\dagger_\ell,
M^{}_{\rm D} M^\dagger_{\rm D} \right]$, which should more or less be
associated with the lepton-flavor-violating decays of charged
leptons \cite{Rode}.
Taking $M^{}_{\rm D} \simeq R \widehat{M}^{}_N U^T_0$, we find
%%%%%%%%%%%%%%%%%%%%%%%%%%%%%%%%%%%%%%%%%%%%%%%%%%%%%%%
\footnote{A much more explicit expression of this quantity is very
lengthy, and hence it will be presented elsewhere \cite{Wang}.}
%%%%%%%%%%%%%%%%%%%%%%%%%%%%%%%%%%%%%%%%%%%%%%%%%%%%%%%
\begin{eqnarray}
{\rm Im}\left(\det\left[ M^{}_\ell M^\dagger_\ell,
M^{}_{\rm D} M^\dagger_{\rm D} \right]\right)
\hspace{-0.2cm} & \simeq & \hspace{-0.2cm}
2\Delta^{}_{e\mu} \Delta^{}_{e\tau} \Delta^{}_{\mu\tau}
{\rm Im}\left[\left(M^2_1 \hat{s}^{}_{14} \hat{s}^{*}_{24} +
M^2_2 \hat{s}^{}_{15} \hat{s}^{*}_{25} +
M^2_3 \hat{s}^{}_{16} \hat{s}^{*}_{26} \right) \right .
\nonumber \\
\hspace{-0.2cm} & & \hspace{2.8cm}
\left(M^2_1 \hat{s}^*_{14} \hat{s}^{}_{34} +
M^2_2 \hat{s}^*_{15} \hat{s}^{}_{35} +
M^2_3 \hat{s}^*_{16} \hat{s}^{}_{36} \right)
\nonumber \\
\hspace{-0.2cm} & & \hspace{2.78cm}
\left. \left(M^2_1 \hat{s}^{}_{24} \hat{s}^{*}_{34} +
M^2_2 \hat{s}^{}_{25} \hat{s}^{*}_{35} +
M^2_3 \hat{s}^{}_{26} \hat{s}^{*}_{36} \right)\right] \; , ~~~~
%     (16)
\end{eqnarray}
which depends on six independent phase differences, such as
$\delta^{}_{2i} - \delta^{}_{1i}$ and $\delta^{}_{3i}-\delta^{}_{1i}$
(for $i=4,5,6$). These phase parameters are apparently different
from those governing ${\cal X}^{}_N$, as one can see from Eq. (11).
It is therefore
desirable to search for $\mu \to e + \gamma$ and other possible
lepton-flavor-violating channels, so as to fully probe the seesaw
mechanism and its parameter space. In the minimal supersymmetric
standard model extended with three heavy Majorana neutrinos, for
example, there is some parameter space for the branching ratios
of $\mu \to e + \gamma$ and $\tau \to \mu + \gamma$ decay modes
to be close to their respective upper bounds as set by the present
experiments \cite{Antusch2}. The next-generation experiments of this
kind are expected to impose more stringent constraints on such rare or
forbidden processes in the SM and on possible new physics behind them.

\framebox{\large\bf 3} \hspace{0.1cm}
At this point it makes sense to comment on the weak-basis invariants
$I^{}_i$ as proposed by Branco {\it et al} \cite{Branco} in the
canonical seesaw mechanism. Now that the several leptonic
commutators discussed above are independent of the flavor basis of
weak interactions taken for the lepton mass matrices, one of them
should be more or less equivalent to $I^{}_i$, which are defined as
\begin{eqnarray}
I^{}_1 \hspace{-0.2cm} & \equiv & \hspace{-0.2cm} \text{Im}\text{Tr}
\left[\left(M^\dagger_{\rm D} M^{}_{\rm D}\right)
\left(M^\dagger_{\rm R} M^{}_{\rm R}\right) M^*_{\rm R}
\left(M^\dagger_{\rm D} M^{}_{\rm D}\right)^* M^{}_{\rm R} \right]
\; , \nonumber \\
I^{}_2 \hspace{-0.2cm} & \equiv & \hspace{-0.2cm} \text{Im}\text{Tr}
\left[\left(M^\dagger_{\rm D} M^{}_{\rm D}\right)
\left(M^\dagger_{\rm R} M^{}_{\rm R}\right)^2 M^*_{\rm R}
\left(M^\dagger_{\rm D} M^{}_{\rm D}\right)^* M^{}_{\rm R} \right]
\; , \nonumber \\
I^{}_3 \hspace{-0.2cm} & \equiv & \hspace{-0.2cm} \text{Im}\text{Tr}
\left[\left(M^\dagger_{\rm D} M^{}_{\rm D}\right)
\left(M^\dagger_{\rm R} M^{}_{\rm R}\right)^2 M^*_{\rm R}
\left(M^\dagger_{\rm D} M^{}_{\rm D}\right)^* M^{}_{\rm R}
\left(M^\dagger_{\rm R} M^{}_{\rm R}\right) \right] \; .
%     (17)
\end{eqnarray}
By construction, these three invariants are only sensitive to the
CP-violating phases which appear in leptogenesis, because $M^{}_{\rm D}$
always appears in the form of $M^\dagger_{\rm D} M^{}_{\rm D}$. So CP
invariance requires $I^{}_1 = I^{}_2 = I^{}_3 = 0$. Given $M^{}_{\rm
D} \simeq R \widehat{M}^{}_N U^T_0$ and $M^{}_{\rm R} \simeq U^{}_0
\widehat{M}^{}_N U^T_0$ in the seesaw approximation, we find that
the structures of $I^{}_i$ (for $i=1,2,3$) are almost the same:
\begin{eqnarray}
I^{}_1 \hspace{-0.2cm} & \simeq & \hspace{-0.2cm} \text{Im}\text{Tr}
\left[\left(R^\dagger R\right) \widehat{M}^5_N \left(R^\dagger R
\right)^* \widehat{M}^3_N \right] = \sum_{i < j} \left(M^{}_i M^{}_j
\right)^3 \left(M^2_j - M^2_i\right) {\rm Im} \left[\left(R^\dagger
R\right)^{}_{ij}\right]^2
\; , \nonumber \\
I^{}_2 \hspace{-0.2cm} & \simeq & \hspace{-0.2cm} \text{Im}\text{Tr}
\left[\left(R^\dagger R\right) \widehat{M}^7_N \left(R^\dagger R
\right)^* \widehat{M}^3_N \right] = \sum_{i < j} \left(M^{}_i M^{}_j
\right)^3 \left(M^4_j - M^4_i\right) {\rm Im} \left[\left(R^\dagger
R\right)^{}_{ij}\right]^2
\; , \nonumber \\
I^{}_3 \hspace{-0.2cm} & \simeq & \hspace{-0.2cm} \text{Im}\text{Tr}
\left[\left(R^\dagger R\right) \widehat{M}^7_N \left(R^\dagger R
\right)^* \widehat{M}^5_N \right] = \sum_{i < j} \left(M^{}_i M^{}_j
\right)^5 \left(M^2_j - M^2_i\right) {\rm Im} \left[\left(R^\dagger
R\right)^{}_{ij}\right]^2\; ,
%     (18)
\end{eqnarray}
where $i,j = 1,2,3$. It becomes obvious that $I^{}_1$, $I^{}_2$ and
$I^{}_3$ contain the same information about CP-violating phases.
Taking $I^{}_1$ for example and taking account of the the explicit
parametrization of $R$ in Eq. (8), we immediately arrive at
\begin{eqnarray}
I^{}_1 \hspace{-0.2cm} & \simeq & \hspace{-0.2cm} \left(M^{}_1
M_2\right)^3 \Delta^{}_{21} \text{Im}\left(\sum^3_{i=1}
\hat{s}^{}_{i4} \hat{s}^*_{i5}\right)^2 + \left(M^{}_1 M_3\right)^3
\Delta^{}_{31} \text{Im}\left(\sum^3_{i=1} \hat{s}^{}_{i4}
\hat{s}^*_{i6}\right)^2 \nonumber \\
& & \hspace{5.3cm} + \left(M^{}_2 M_3\right)^3 \Delta^{}_{32}
\text{Im}\left(\sum^3_{i=1} \hat{s}^{}_{i5} \hat{s}^*_{i6}\right)^2
\; .
%     (19)
\end{eqnarray}
We see that $I^{}_1$ depends on the same (six independent)
CP-violating phases as ${\cal X}^{}_N$ does, but they are not equivalent
to each other. In fact, the CP-violating asymmetries $\varepsilon^{}_{i\alpha}$
or $\varepsilon^{}_i$ vanish if all the three phase terms in Eq. (19)
vanish, or equivalently the three invariants $I^{}_1$, $I^{}_2$ and
$I^{}_3$ are all vanishing \cite{Branco}.

\framebox{\large\bf 4} \hspace{0.1cm} In summary, we have constructed
the commutators of lepton mass matrices and explored their relations
with CP violation in the canonical seesaw and leptogenesis mechanisms.
It is demonstrated that ${\rm Im}\left(\det\left[ M^\dagger_{\rm D} M^{}_{\rm D},
M^\dagger_{\rm R} M^{}_{\rm R} \right]\right)$, ${\rm
Im}\left(\det\left[ M^{}_\ell M^\dagger_\ell, M^{}_\nu M^\dagger_\nu
\right]\right)$ and ${\rm Im}\left(\det\left[ M^{}_\ell
M^\dagger_\ell, M^{}_{\rm D} M^\dagger_{\rm D} \right]\right)$ can
serve for a basis-independent measure of CP violation associated
with the lepton-number-violating decays of heavy Majorana neutrinos, the
flavor oscillations of light Majorana neutrinos and the
lepton-flavor-violating decays of charged leptons, respectively. We
have calculated these rephasing-invariant quantities with the help of a
standard parametrization of the $6\times 6$ flavor mixing matrix,
and discussed their implications on both leptogenesis and CP violation
at low energy scales. We have also calculated the weak-basis invariants
of leptogenesis as defined by Branco {\it et al} \cite{Branco} and
pointed out their similarity with and difference from 
${\rm Im}\left(\det\left[ M^\dagger_{\rm D} M^{}_{\rm
D}, M^\dagger_{\rm R} M^{}_{\rm R} \right]\right)$ in our case.

Finally, let us emphasize that the {\it commutator} language has played a very
important role in the developments of Quantum Mechanics and Quantum Field
Theories, and its applications in flavor physics have proved to be
interesting and instructive for a basis-independent description of
flavor mixing and CP violation. For example, the leptonic commutator
$\left[ M^{}_\ell M^\dagger_\ell, M^{}_\nu M^\dagger_\nu \right]$ in
vacuum and its counterpart in matter can help establish some direct
relations between the effects of CP and T violation in vacuum and those in matter
\cite{Xing2001,HS,JX}. This kind of study is therefore useful for the
upcoming long-baseline neutrino oscillation experiments. The present work,
which has offered a novel application of the {\it commutator} language in
the canonical seesaw and leptogenesis mechanisms, is also meaningful and
helpful to enrich the phenomenology of neutrino physics.

\vspace{0.5cm}

One of us (Y.K.W.) would like to thank the Theoretical Physics
Division of IHEP, where this work was done, for hospitality and
support. The research of Z.Z.X. was supported in part by the
National Natural Science Foundation of China under grant No.
11135009.

\end{document}